\documentclass[11pt]{article}
\usepackage{moriond,epsfig}

\def\be{\begin{equation}}
\def\ee{\end{equation}}
\def\bea{\begin{eqnarray}}
\def\eea{\end{eqnarray}}

\begin{document}
\vspace*{3cm}
\title{ORIGINS OF QUASARS AND GALAXY CLUSTERS}

\author{ H. ARP }

\address{Max-Planck-Institut f\"ur Astrophysik, 85741 Garching, Germany}

\maketitle\abstracts{
The distribution on the sky of clusters of galaxies shows significant association with 
relatively nearby, large, active galaxies. The pattern is that of clusters paired equidistant
across a central galaxy with the apparent magnitudes and redshifts of their constituent 
galaxies being closely matched. The clusters and the galaxies in them tend to be strong
X-ray and radio emitters and their redshifts occur at preferred redshift values. {\it The 
central, low redshift galaxies often show evidence of ejection in the direction of these 
higher redshift clusters and the clusters often show elongation along these lines.} 
In most of these respects the clusters resemble quasars which have been
increasingly shown for the last 34 years to be similarly associated with active parent
galaxies.  It is argued here that, 
empirically, the quasars are ejected from active galaxies. They evolve to lower redshift 
with time, fragmenting at the end of their development into clusters of 
low luminosity galaxies. The cluster galaxies  can be at the same distance as their lower 
redshift parents because they still retain a component of their earlier, quasar intrinsic 
redshift. }

\section{Inroduction}

{\bf The distribution on the sky of clusters of galaxies started to be catalogued about 40 
years ago by George Abell and collaborators. The cores of these clusters were predominantly old
stellar population E galaxies which were believed to be mostly gas free and inactive.
With the advent of X-ray surveys, however, it became evident that many clusters of
galaxies were strong X-ray emitters. This evidence for non-equilibrium behavior was
not easily explained. In these active properties, however, the clusters joined AGN's and
quasars as the three principal kinds of extragalactic X-ray sources. Evidence then
developed that quasars, and now some galaxy clusters were physically associated with much 
lower redshift galaxies.  Surprisingly, the cluster redshifts were sharply peaked at the preferred 
quasar redshifts of z = .061, .30 etc. (This evidence has been discussed principally in Arp 1997; 1998a; 
Arp and Russell 2001).}

It was possible to explore these properties further by plotting the distribution of 
galaxy clusters on other, larger areas on the sky. Some appeared projected along the spine of 
the Virgo Cluster. It turned out that the  
Abell clusters  which were located in that part of the sky in the direction of Fornax fell 
in the same distinctively elongated area as the large, low redshift Fornax Cluster. (The 
Abell clusters reach to about z = .2  limit and the brightest galaxy in the Fornax cluster 
is z =.0025.) On the sky, in the direction of the giant, low redshift galaxy CenA/NGC5128, the Abell 
clusters fell almost exclusively along a broadening extension of the X-ray, radio jet going northward 
from this active galaxy. This is the same line occupied by a number of active, higher 
redshift galaxies which have been previously associated with ejection of radio plasma 
from CenA (Arp 1998a)

\section{Abell Clusters A3667 and A3651}

   Abell 3667 is a rich cluster of galaxies studied in radio and X-rays by Rottgering
et al. (1997) and Knapp, Henry and Briel (1996). It emits copious X-rays. (Grandi et al.
1999 estimate  a total of 2,440 ct/ks in the hard band.) Its galaxies are bright, with the 
tenth brightest having an apparent magnitude of $m_{10}$ = 15.3 mag.  As Fig.1
shows, there is only one other rich cluster of bright galaxies for a wide area
around. That second cluster is A3651, a cluster with almost identically bright galaxies, 
$m_{10}$ = 15.4 mag.  A3651 also is a bright X-ray emitter with a total of 430 ct/ks.
The two form an obvious pair of galaxy clusters.

\begin{figure}[h]
\psfig{figure=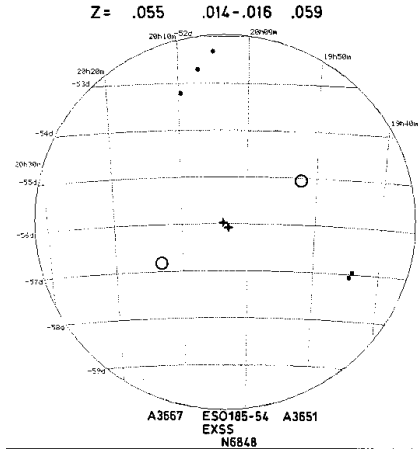,height=6cm}
\hfill \parbox[b]{6.0cm}{\caption{A particularly conspicuous pair of bright galaxy clusters with
closely the same redshift is shown. Abell clusters with m$_{10} \leq$ 15.4 mag are plotted as open 
circles.The remaining clusters in the field are designated as small filled circles and have m$_{10}
\geq$ 17.3 mag. The plus signs indicate the two brightest galaxies in a group with redshift
near z = .015. The two paired clusters have redshifts near z =.06.}}
\label{moriondfig1}
\end{figure}

     As Fig. 2 shows, A3667 is a very elongated cluster and points directly at A3651.
The latter, in turn, is elongated back toward A3667. Almost exactly at the center
between these two clusters are located some very bright galaxies.

\begin{figure}[h]
\psfig{figure=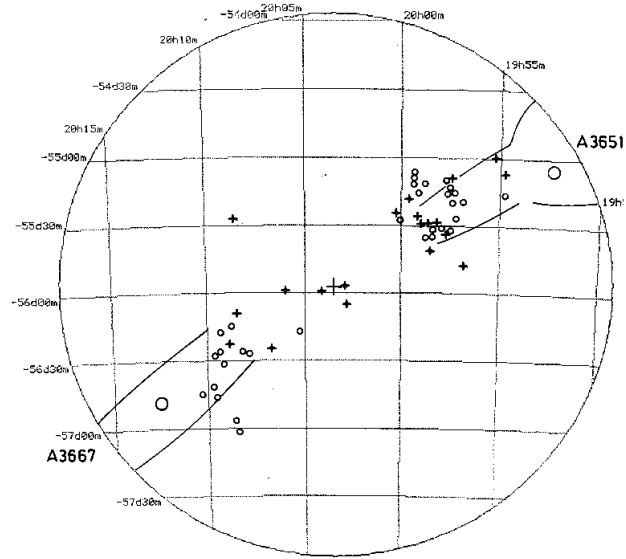,height=8cm}
\hfill \parbox[b]{6.0cm}{\caption[]{An enlarged plot of Fig. 1 shows the two clusters Abell 3667 and 3651 
elongated toward the central group. The shape of these two clusters is indicated by
boundary lines and their centers by large circles. On the inner edges individual galaxies 
belonging to the z = .06 clusters are indicated as small open circles. Individual galaxies 
belonging to the z = .015 group are shown as plus signs.

The remarkable result which appears is that the low redshift galaxies reach out from the center and
actually intermingle with the higher redshift galaxies of both clusters}}
\label{moriondfig2}
\end{figure}

For example
NGC6848  is 13.1 mag. An even brighter galaxy, however, which surprisingly, was 
missed by early cataloguers is ES0 185-54 at $m_{B}$ = 11.9 mag. The latter is the
dominant galaxy in a group of galaxies which is called MdCL 15 (Maia et al. 1989).
Ramella, Focardi and Geller (1996) give a mean redshift of z = .0157 for this group.

     The most startling result, however, is shown in Fig. 2 where the galaxies of redshift 
near 5,000 km/sec extend out on both sides of the central group and join and intermingle
with the galaxies of around 17,000 km/sec. The configuration is striking even when all
the galaxies in the area are plotted (e.g. in SIMBAD) regardless of whether their 
redshifts are known or not. {\it The two elongated X-ray clusters appear to be
continuous, linear extensions of the bright galaxies surrounding the central ESO
185-54 galaxy.}

    Such a coincidence seems extremely unlikely to occur by chance. Since there 
is a strong precedent for pairing of higher redshift objects such as quasars and
higher redshift companions across active galaxies, it is natural to ask whether the
the low redshift galaxy in the center is active. It turns out that ESO 185-54 is an
early type galaxy with bright emission lines (da Costa et al. 1988: Sairava, Ferrari and
Pastoriza 1999). 

By good fortune the field was also
observed for 5752 seconds with the ROSAT PSCP because of the presence of a white
dwarf. That observation shows that ESO 185-54 is an extended X-ray source of
about 27 ct/ks (broad band). Reduction of that exposure 
reveals that there is indeed a line of bright X-ray sources across the central X-ray
galaxy, NW and SE, closely along the line to either of the two strong X-ray clusters.
These sources are respectively double, triple and double. X-ray sources ejected 
from active galaxies are often double or triple (Arp 2001) suggesting that these are incipient 
quasars in the process of fragmenting and evolving into
groups and clusters of lower redshift galaxies. 

      It can also be seen in the full PSPC broad band exposure that 
the majority of the fainter X-ray sources in both the central group around ESO 185-54 and the rich 
X-ray cluster A3667 are elongated toward each other in the same way the galaxies are in Fig. 2. 
The lines of X-ray sources 1, 2 and 3 in Fig. 5 are at p.a. = 137 and 
305 deg. whereas the directions to the X-ray clusters are at about 125 and 300 deg. respectively.
There is also a high redshift cluster of z=.710 along the line to A3667. Further details are given in Arp
and Russell 2001.

 {\bf Note Added.} After the submission of this manuscript a preprint appeared by
Vikhlinin, Markevitch and Murray (2001) reporting that Chandra observations of Abell 
3667 showed a bow shock indicating that it was moving through the intergalactic 
medium with a speed of about 1400 km/sec. They state "The edge is . . . almost
perpendicular to the line connecting subclusters A and B." that would place it at p.a. 
= 122 to 127 degrees. But it has just been noted in the closing paragraph above
that the inferred line of SE ejection from the central ESO 185-54 is about p.a. = 125 deg.
{\bf  We therefore now have direct evidence for Abell 3667 moving accurately out along 
this line of ejection which had been previously predicted.}

\section{Recent Results on the Surroundings of M101}

In addition to the case discussed above, the recent investigation by Arp and Russell (2001) reported a 
number of other cases of galaxy clusters paired 
across large, nearby galaxies. For example, an elongated cluster Abell 2256 pointing toward NGC 6217
(Atlas of Peculiar Galaxies No. 185) with quasars of z = .380, .376 and .358 closely around this X-ray 
jet galaxy. A pair of 3C radio quasars across another Atlas object, No. 227 had a probabiliy of only
about $2x10^{-9}$ of being accidental. This pair defined an ejection direction which ended on a total of
11 bright Abell clusters. A number of other cases were presented which had colinear strings of 
quasars and galaxy clusters emerging from large, low redshift galaxies which had negligible chance of 
being conicidences.
 
One of the cases presented in the above reference was that of two bright clusters of z = .070 and 
z = .071 aligned diameterically across the nearby, bright apparent magnitude, ScI spiral, M101. 
I had known that Markarian 273, regarded 
as one of the extreme "ultra luminous" infrared galaxies, was relatively close to M101 on the side 
toward the z = .070 galaxy cluster. But then I happened to notice the Hickson Compact Group No. 66 
at z = .070. It turned out to be in the direction of M101 from the cluster with z =.070. While considering 
this development, an astronomer with a wider overall awareness, Amy Acheson, sent me a question: 
"Why do the famous, active objects like 3C295 tend to fall close to bright galaxies?" I plotted the
position of 3C295 as in Fig. 1 here and therepon decided that I had to investigate in detail all the 
various kinds of objects around this particularly large, nearby galaxy.

\begin{figure}
\psfig{figure=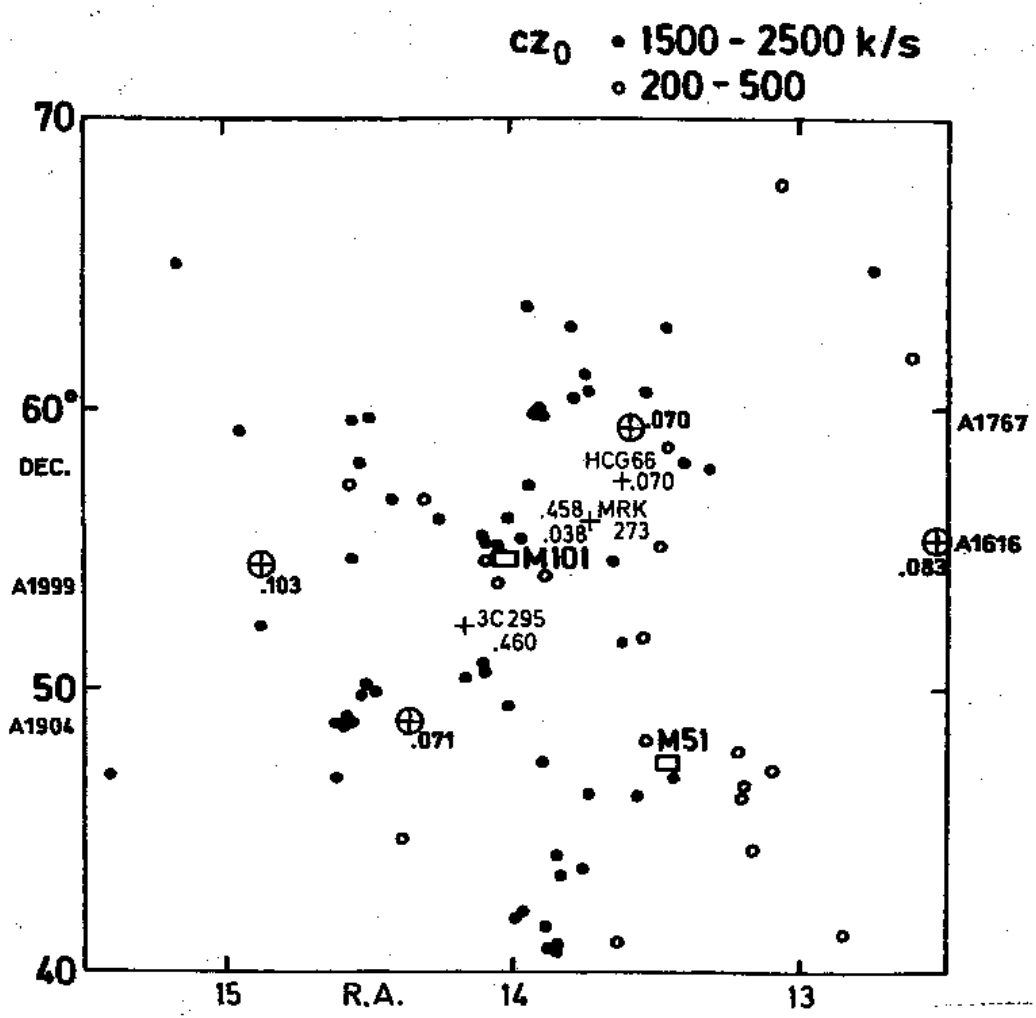,height=13.7cm}
\caption{Over a wide area are plotted companions around the nearby galaxy, M101 
(200 - 500 km/sec). Also galaxies of redshift 1500 - 2500 km/sec appear concentrated 
around M101, some in a predominantly NW - SE line. The circled plus signs represent all 
Abell clusters with m$_{10}$ brighter than 16.3 mag. The clusters which pair NW -SE 
across M101 have mean redshifts of z = .070 and .071. The plus signs designate special
objects described in text.
\label{M101fig1}}
\end{figure}

The filled circles in Fig. 1 show a plot of all galaxies between cz = 1500 and 2500 km/sec over a large 
area around M101. They are very sparse on the edges of the region but increase steadily toward the 
center, to the position where M101 is located. It is difficult to see how this concentration toward
M101 could be a selection effect since galaxies of both higher and lower redshift are scattered more 
or less uniformly throughout the area. (See Arp 1990 for details of this analysis.) Moreover, these 
galaxies form lines radiating from the position of M101 outwards. The line radiating NW - SE from 
M101, however,  is the most populated. The open circles show low redshift companions of M101 
reaching out into the same areas - demonstrating that physical association with M101 reaches out to 
this distance. 

The latest update in Fig. 1, shown here, has marked the positions of the brightest Abell Clusters (As 
in the earlier Fig. 7 of Arp and Russell 2001). But also added are three plus signs showing the positions 
of Mrk 273, HCG 66 and 3C295. These three objects, along with two bright Abell clusters fall roughly in 
the filament of galaxies which extend in the NW - SE direction which was identified in 1990.

Mrk 273 is one of the three major, ultra luminous infrared galaxies known (Arp 2001). It is remarkable
that it falls this close to, and with this orientation with respect to M101. It is a strong X-ray and radio
source and in this respect is similar to higher redshift, active objects ejected from more distant
central galaxies. In the present case, however, the apparent brightness of both the central galaxy and 
the AGN imply a closer distance and account for the larger apparent separation ( $\sim$ 3 deg.).

Further along in this direction we encounter HCG 66 (Hickson 1994). This is a compact chain of six
galaxies four of which have redshifts which average to z = .070. As a group it is like a small,  or sub 
cluster (albeit in a non equilibrium configuration). With its identical redshift it would seem to be related
to Abell 1767 just to the NNW. In this respect it would seem to be an extension of Abell 1767 in the
general direction of the filament of objects leading back to M101. This will be an important property, of 
the galaxy clusters which appear to be involved in the alignments with the central galaxy. More 
elongations of clusters along the line back to their galaxy of origin will be shown later in this paper. As 
for the long apparent extension on the sky across M01 (see also Arp 1984), the only longer line of 
apparent separation known is the ls the line of higher redshift galaxies extending along the minor axis 
of M31 in the Local Group of galaxies, which is obviously much closer to us (see Arp 1998b).

\subsection{3C295}
To the SW of M101 in Fig. 1 is plotted the famous radio galaxy 3C295. The remarkable properties
of this location are that it is closely the same separation and on the other side of M101 from Mrk 273.
Most striking, however, it is almost exactly on a line to the SE Abell cluster, at z =.071. I
remember the excitement at Mt. Wilson and Palomar observatories when the redshift of 3C295 was
first measured. Among the strongest radio sources discovered in initial radio surveys, Minkowski's 
1960 redshift of z = 0.46 remained the highest redshift measured until past 1975 
(see Sandage 1999 for history of major events over 50 years at Palomar).

The most recent observation of 3C295 is in X-rays by Chandra (Harris et al. 2000). It shows a pair 
of X-ray condensations coming out of the nucleus at a position angle of about p.a. = 144 deg. This is 
the same alignment as the radio lobes shown as overlayed contours. Since the ejection origin of
radio lobes has long been accepted, and X-ray jets often are found at their core, the X-ray
sources in 3C295 are indicated to be in the process of ejection. Suprisingly, at a very small scale, 
between 3C295 at z = .461 and its closest major companion at z =.467, there is a much brighter galaxy 
at z = .285. The two higher redshift galaxies are almost perfectly aligned across the lower redshift 
galaxy at only 11 arcsec on either side of it. There is a only very small a priori chance of a galaxy this 
bright accidently occurring at this exact spot. This of course is the quintessential pattern of AGN's 
ejected from a larger galaxy (often interpreted as gravitational lensing). There is some hint of 
luminous material connecting from the central galaxy to 3C295 and deeper, high resolution images 
should be obtained to check the possibility that this brighter, low redshift galaxy is the origin of 3C295
and its similar redshift companion.

\pagebreak

\section{The Distribution of Bright Markarian Galaxies}

Because of the provocative location of Mrk 273 it seemed interesting to see if there were similar 
objects within the extended field of M101. A Simbad screen was set up to tabulate objects brighter
than about 15.5 mag. with redshifts between 0.35 and 0.41. A field of radius 10 degrees was
arbitrarily set. Fig. 4 shows that a total of five Markarian galaxies were found. Unexpectedly the 
exceptionally active Mark 231 showed up near the NW edge of the field. Mrk 231 is the second of
what are believed to be the three most luminous infrared galaxies in the sky (Arp 2001). To find it
here, close to Mrk 273, and in the same NW direction from M101 appears by chance to be quite 
unlikely.

\begin{figure}[h]
\psfig{figure=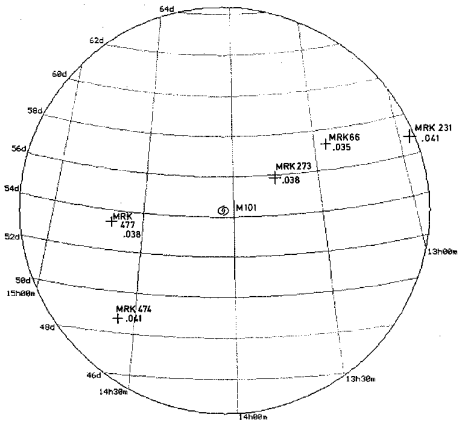,height=6cm}
\hfill \parbox[b]{6.0cm}{\caption{Around the very bright nearby galaxy M101 it shown that five of
the brightest and most intensively studied Markarian (active) galaxies are strung out in a
direction rougly NW - SE. Within a radius of 10 deg. All Markarian galaxies brighter than 15.3 
magnitude and z between .035 and .045 are plotted.}}
\label{M101fig4}
\end{figure}

The second point Fig. 4 illustrates is that the five Markarian galaxies are distributed roughly along a
line NW - SE which, as we shall continue to see, is where most of the galaxies and active objects are
situated.  But these Markarian galaxies are all exceptional objects. The intensity with which they have
been studied attests to their importance - for example the number of literature references for Mrk 231
is 509, for Mrk 273, 274; for Mrk 477, 126; for Mrk 474, 60 and even for the least, Mrk 66, there are 32
references. 

Perhaps even more impressive, however, is that this is a general alignment over almost 20 degrees in
the sky of exceptionally active objects with nearly the same redshift: z = .041, .041, .038, .038, .035.
In terms of conventional redshift distance this would represent a narrow filament of physically related 
galaxies spanning an enormous distance in space. The tendency would be to say that these were all
galaxies at one particular time in their evolution. Associated with M101, however,  they could have an 
origin in a single ejection event. Such events are recurrent and could later furnish different lines of 
objects at different discrete stages in their evolution.    

\section{The Distribution of Bright Quasars}

In order to separate background quasars from candidates for association with M101 another Simbad 
screen was set at the relatively bright apparent magnitude of V = 17.1 mag. The search was
supplemented by visual search of the Veron and Veron Catalog and checks with NED lists of
high redshift quasars. The quasars found are plotted in Fig. 5. The first impression is that these
brightest quasars are distributed along the same general line of Markarian galaxies as just discussed. 
In detail, there are quasars near each of the plus signs which represent the Markarian galaxies from 
Fig. 4, suggesting that these quasars may have originated more recently from these lower redshift, 
active Markarian galaxies and not necessarily as direct ejections from M101 itself.

\begin{figure}[h]
\psfig{figure=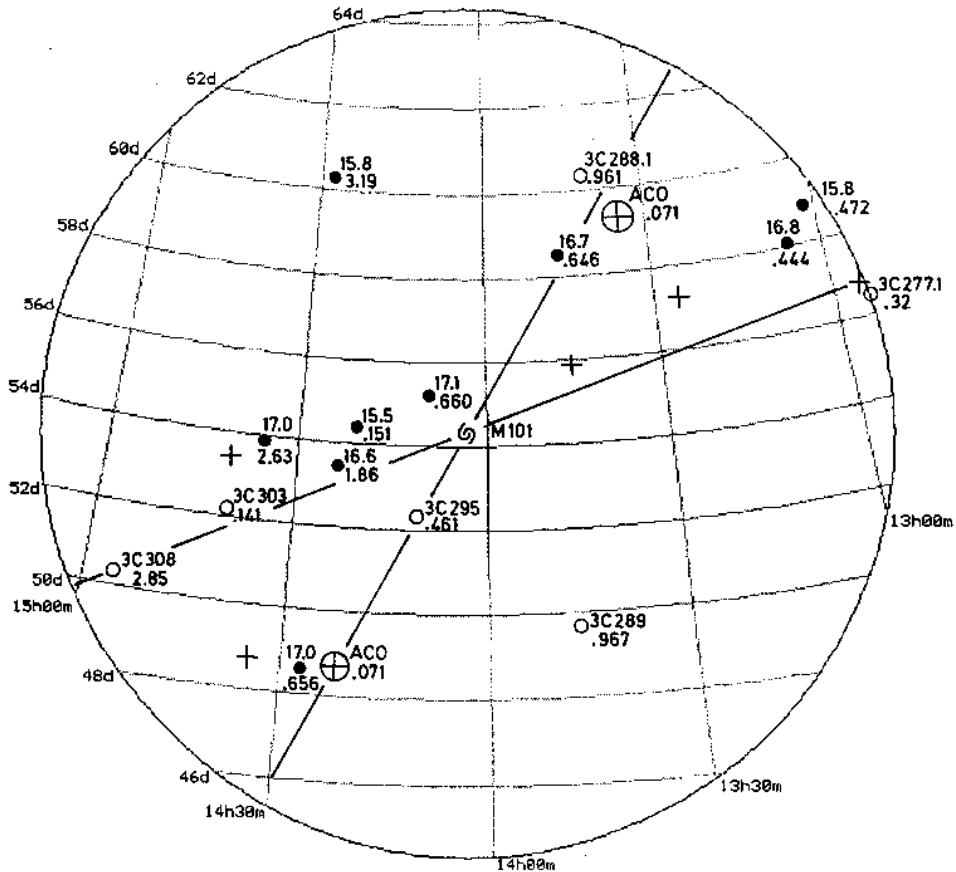,height=5in}
\caption{The plus signs are the same Markarian galaxies from Fig. 4 with now all QSO's less 
than z = 17.1 mag added as filled circles. All known 3C radio objects in the area are represented by
open circles. The two circled plus signs are Abell Clusters from Fig. 1. 
\medskip
\label{M101fig5}}
\end{figure}

Some of the high redshift quasars here have such bright apparent magnitudes that whatever 
the average luminosity for this redshift may be, they are certainly among the closest of this class - if 
the red shift does not indicate distance (Arp 1999). Examples are redshifts of z = 2.63 at 17.0 mag.,
 z = 1.86 at 16.6 mag. and z = 3.19 at 15.8 mag. (We should remark that while the z = 3.19 quasar is 
not near a Markarian galaxy it is quite near an infrared, IRAS, galaxy of 15.3 mag. and z = .037 and 
therefore similar to the Markarian galaxies plotted here.) The 15.5 mag. object is a BL Lac type 
quasar, OQ 530, very bright in apparent magnitude for its class and agreeing with the close 
association of BL Lac's with nearby galaxies as reported in Arp (1997, Table 2 and 1998a).

\subsection{Numerical Coincidences of Redshifts}

 There are some remarkable numerical agreements in redshifts of
a number of objects in Fig. 5. The quasars at z = .646, .660 and .656 stand out. It should be remarked
that there is an additional quasar at 17.7mag., slightly below the cut off of 17.1 mag., which has a 
z= .646 and falls just SW of M101. The latter  makes a close apparent pair with the z = .660 quasar 
across M101 directly. As remarked in the previous section Quasars at greater distances from M101 
may arise from secondary ejections from earlier ejected, still active galaxies such as the Markarian 
objects. 

The numerical coincidence of the 3C quasars at z =.961 and .967 also stands out. Even more strikingly, 
these redshifts agree very closely with peak values of redshifts in the Karlsson formula which 
expresses the empirical relation found for many years for quasar redshifts:
\medskip
                      $$ z = .06, .30, .60, .96, 1.41, 1.96, 2.64, 3.48 . . .$$ 
\medskip
The agreement between these predicted values and the values for the majority of quasars in Fig. 5 is
evident. Recently Burbidge and Napier (2001) have demonstrated a very significant extension of the
Karlsson series to the highest redshifts.

\section{The 3C Radio Sources}

Again in Fig. 5 we see the brightest, earliest discovered, 3C  radio sources falling mostly along the line 
of objects which passes through M101 from NW to SE. Apparently 3C277.1  is associated with Mrk 231
and 3C295 apparently with M101 itself. Some of the others may also be associated with nearby active
galaxies such as Mrk 477.

3C303 is a particularly well studied object because of it has a prominent one sided radio jet pointing
from a compact radio galaxy of z = .141 toward its double radio lobe, the northern component
of which is at p.a. = 280 deg. (Lonsdale et al. 1983).The direction to M101 is about 289 deg. The 
ROSAT PSPC observation shows X-ray sources mildy extended in the NW -SE direction. 

Of particular importance, however, are three ultraviolet excess objects apparently
associated with the western lobes of 3C303 (Kronberg et al. 1977). Margaret Burbidge obtained 
from the spectrum of one a redshift of z = 1.57. In spite of the fact that this association was 
discussed as a possible conclusive proof of the near distance of quasars, no further time has 
been assigned to obtain the spectra of the remaining two candidates. 

With the plot in Fig. 5 now complete, we could propose that there was a cone of ejected objects from 
M101 in the NW - SE direction.  Or we could interpret the distribution of objects as two rather narrow 
lines, one at p. a. = 110 and the other at p.a. = 147 deg. We will prefer the latter interpretation because
we now wish to examine the possible connection of the Abell galaxy clusters with these lines of 
objects. 

\section{Elongation of Clusters Toward M101}

From previous evidence we have some expectation of galaxy clusters being elongated along the line of
their apparent ejection. In Arp and Russell (2001) the clusters Abell 3667 and 3651 were paired across 
the central galaxy ESO 185-54. The clusters were strongly elongated along this connecting line and 
later Chandra observations actually showed a bow shock indicating motion outward along just this line 
at 1400 km/sec. The famous gravitational arc cluster, Abell 370 showed elongation back toward its 
purported galaxy of origin, the bright Seyfert NGC1068. Several other similar examples were also 
noted in the above reference.
 
Cataloged galaxies within a radius of 60 arcmin around the cluster Abell 1904 show that he
cluster is noticeably elongated in a direction leading back to 3C295, and hence very closely along the
line back to M101. Another cluster of somewhat fainter galaxies, Abell 1738, which is just N of Mrk 66 
has relatively few cluster members with measured redshifts but they are aligned very accurately 
back toward M101. 

Not much can be said about Abell 1767 except that it gives some impression of being aligned back 
toward M101 at about 140 deg. The PSPSC X-ray sources around 3C303 appear inclined at about 
p.a. = 110 deg toward M101. These last two clusters, however, would require quantitative 
determination of their optical galaxy isopleths and/or X-ray countours of their smoothed distribution.

The shapes of the clusters, however, in conjunction with similar data 
referenced, would seem to furnish compelling evidence for the origination of the cluster from a point
near the present M101. In the conventional view, constraints which give spontaneuosly forming, 
linear, non equilibrium configurations in deep space would seem to difficult enough. But to have them 
pointing back to a large galaxy would seem to require their connection with generally linear, 
recurrent ejections from that galaxy.

\subsection{An Additional Example of Aligned, Connected Clusters of Different Redshift }

\begin{figure}[h]
\psfig{figure=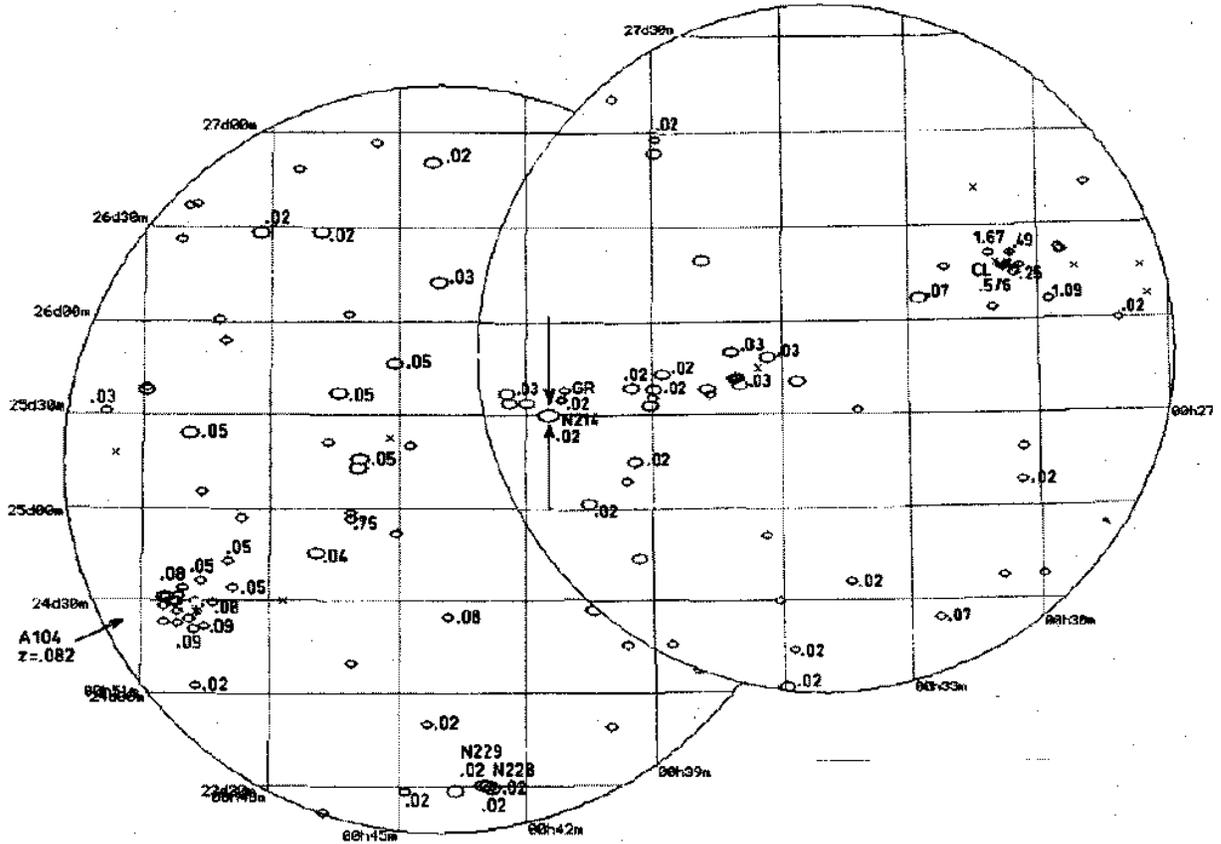,height=12cm}
\caption{All catalogued galaxies from SIMBAD centered on the ScI galaxy 
NGC 214. Known redshifts are labeled. The redshifts increase as the line extends toward the galaxy 
cluster at z = .516 about 2.5 degrees to the WNW and also in the opposite direction about 2 degrees 
toward the Abell cluster 104 at z =.082.
\label{M101fig9}}
\end{figure}

Fig. 6 shows a serendipitously discovered association which well illustrates the elongation of clusters
back toward a central galaxy. It was discovered because I was interested in the unusual galaxy
cluster CRSS J0030.5 +2618 . The cluster has z = .516 and an order of magnitude excess of Chandra 
X-ray sources (Brandt et al. 2000). It also has two galaxies of redshift z = .247, one of z = .269 and 
three quasars of z = .492, 1.665 and 1.372 all within 5 arcmin and a further quasar z = 1.094 within 8
arc min. 

I asked the question : "Where is the bright galaxy from which all this was ejected?" I found it about 2.5
degrees away. It was NGC 214, an ScI galaxy of $B^{o,i}_T$ = 12.48 mag. and a redshift of 
$v_o = 4757$ km/sec or z = .016. This redshift indicates it is associated with the ubiquitous 
Perseus-Pisces filament (Arp 1990). But he striking feature was that bright companion galaxies at 
z = .02 and .03 stretched away to the WNW directly toward the cluster of high red shift objects 
centered around the CRSS cluster. 

Naturally I then asked: "What is on the other side of NGC 214?" As Fig. 6 shows there is an Abell
Cluster, ACO 104 about 2 degrees away and nearly opposite NGC 214 with a red shift of z = .082. The 
striking feature of this cluster is that it is clearly elongated back toward NGC 214. Also interesting is 
the fact that galaxies of z = .04 and z = .05 lead into the cluster from the direction of NGC 214. (There 
are even two galaxies of z = .03 seen inside the cluster at higher resolution.) 

In all these cases all the galaxies and optical objects in these areas should be completely measured.
But even at this stage it seems clear that there are strings of objects extending in opposite directions
from the central galaxy NGC 214 {\it and their redshifts continually increase as we approach galaxy
clusters at either terminus.}

\section{Sunyaev-Zeldovich Effect}

The calculations of the distances of galaxy clusters from their scattering of microwave background in
conjunction with measurements of their X-ray surface brightness seems to rest on such proven 
physical principles that it is difficult to see how anyone could accept much closer distances as the
observations in the present paper claim. If we are, however, to give the observations even equal
consideration we must face the apparent discordance of nearby galaxy clusters with S-Z distance
determinations.

Perhaps the first point to be made is that with galaxy clusters which are nearby we would be dealing 
with strongly
non-equilibrium physical conditions. If they are formed in more recent ejections from active galaxies
and they themselves are ejecting secondarily and in non equilibrium configurations, perhaps it is
incorrect to assume equilibrium temperature, radiation and energy densities. A strong indication of this
situation comes from consideration of cooling flows. In many clusters the densities near the center
are so high that the cooling flows would exhaust the available heat in much less than a cosmic time
scale. The situation is so severe that suggestions have been made for merging or accretion of
companion objects to resupply the energy in the center (A. Fabian, Moriond Conference).

Actually the situation in many clusters, especially those dominated by a large central galaxy, is that
powerful ejections take place which must intermittently add bursts of energy into the surrounding
cluster medium. (See for e.g. 3C295 pictured in Harris et al. 2000) The question then becomes: Do
observations of the cluster medium at any given time represent an equilibrium physical state on which
the conventional physical calculations can be made?

To make this suggestion more specific, consider that in order to account for the intrinsic redshifts of
the objects of various ages associated with an active galaxy it has been necessary to make a more
general solution of the general relativistic field equations where particle masses are a function of time
(Narlikar and Arp 1993). In the beginning this requires a plasma of low mass particles to be ejected
near light velocity. Because of their low mass the newly ejected particles can have large
scattering cross sections, enabling microwave photons to be efficiently boosted. The 
synchrotron/bremstrahlung jets however are observationally well collimated wich must mean
relatively low temperatures
orthogonal to the direction of travel. Their energy is mostly converted to temperature only at the 
interaction surface of the jet cocoon with the surrounding medium.  Both of these, the strong 
scattering and low temperature  effects would tend to give large distances in the SZ equation for 
distances of a small total mass cluster. 

Put in a possibly observable way, even if there were a nearby cluster in temporary temperature 
equilibrium with a well determined X-ray surface brightness, but no measurable SZ microwave 
depression, would there be an upper limit calculated for the distance of this cluster?

\pagebreak

\section{An On-site Demonstration of the Origin of Galaxies}

A recently announced observational result on the ultra luminous infrared galaxy Arp
220 serves to illustrate the formation origin of groups and clusters of galaxies. This ULIRG is reputed
to be one of the most luminous galaxies known and extremely active both optically and in X-rays. 
Investigation of the brighter X-ray sources immediately around the galaxy have identified quasars and
quasar candidates (Arp et al. 2001). Most striking are a pair of quasars shown here in Fig. 7 which are
exactly aligned across the central active galaxy. By now many such pairs have been identified across
active galaxies and they tend to have similar redshifts. But the striking result in Fig. 7 is that this pair
have almost identical redshifts of z = 1.26 and 1.25. 

\begin{figure}[h]
\psfig{figure=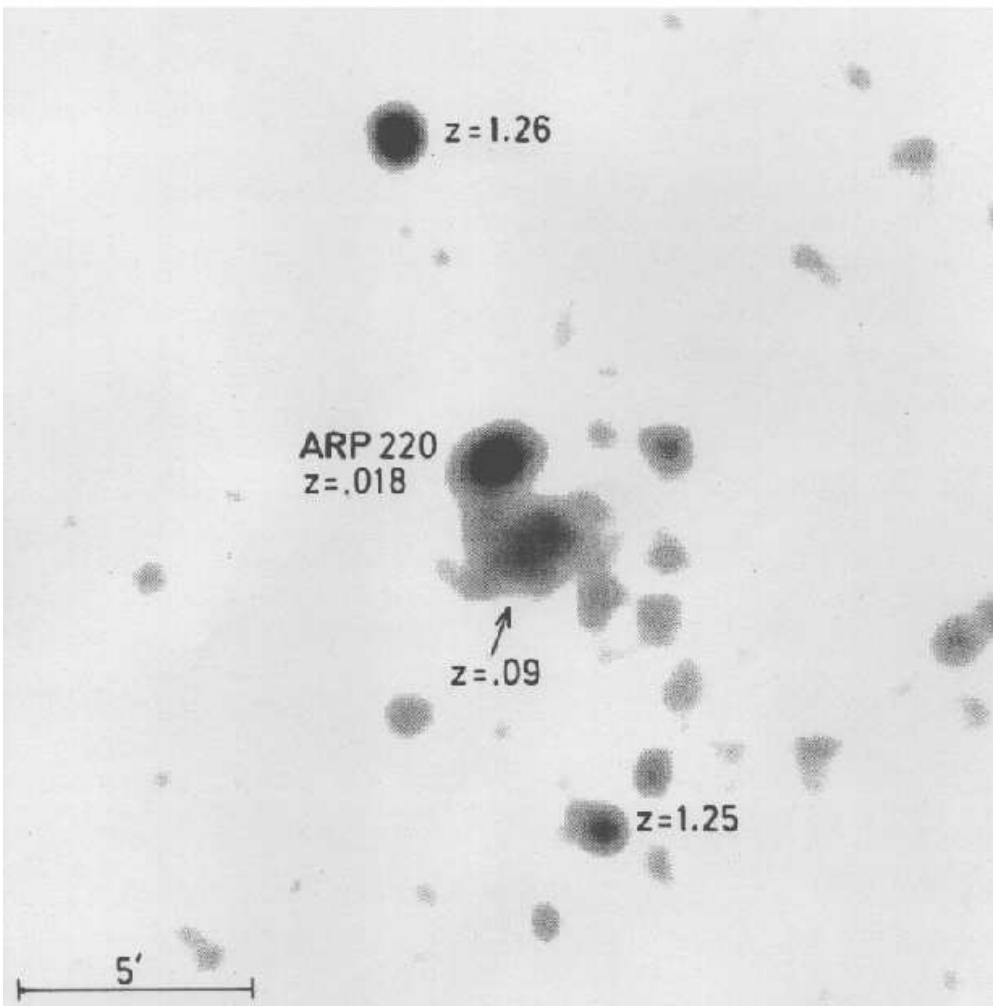,height=17cm}
\caption{Hard X-ray band (.5 to 2.4 keV) from ROSAT PSPC, showing pair of strong sources across
Arp 220. Note curved string of sources leading down to the SSW quasar. Known redshifts are labeled.
\label{moriondf7}}
\end{figure}

Normally such a pair would differ from their mean by of the order of 0.1 in redshift, representing an 
ejection velocity toward and away from the observer from about 10,000 to 30,000 km/sec. There are 
too many close matches in redshift, however to be explained by an ejection origin which
was occasionally closely across the line of sight. These very close matches are more plausibly 
explained by the interaction of the ejected proto quasars with the material in the ejecting galaxy and 
the immediately surrounding medium. This robs the ejecta of their outward velocity, stops them close
to their galaxy of origin and leaves only the intrinsic redshift to be observed (Arp 1999).
Interaction with the body of Arp 220 would furnish a natural explanation for the spectacular disruption 
of this galaxy. 

But the pertinence of Fig. 7 for the discussion in the present paper is that there are three galaxies
emerging from the SW end of Arp 220, all three of which have redshifts about z = .09. These galaxies
are fairly normal looking but are connected back to Arp 220 by both radio (see Arp 2001) and X-ray
material as can be seen so clearly in the Figure. They are strong X-ray sources on their own and
clearly cannot be unconnected background galaxies. In fact they are so bright in apparent magnitude
that if they were at their conventionally believed redshift distance they would have the supposed
luminosities of quasars. They instead appear to be quasars that were severly impeded on their exit
from Arp 220, broke into separate pieces and evolved to their present size and redshift while still in
the edge of the parent galaxy. Unlike most of the cases we have seen where the quasars do not
evolve into lower redshift galaxies until they are at extreme separations from their origin, the present
group represents the rarer case where they have been trapped near the galaxy that they have
disrupted.

This picture is further reinforced by the trail of X-ray sources coming out of Arp 220 to the SSW
quasar. This is on the side of the exit of the z = .09 galaxies and suggests that the trail of X-ray
sources represents material stripped off the fainter of the two quasars as it passed through the region
where the incipient cluster of galaxies is emerging. Regardless of the details, however, this image
seems to show directly from observation how quasars, galaxy groups and galaxy clusters originate
from nearby, low redshift, active galaxies.

\medskip

\centerline{\bf References}

\bigskip

\noindent Arp, H. 1966, {\it Atlas of Peculiar Galaxies},
California Institute of Technology; ApJS 123, Vol. XIV
\medskip

\noindent Arp H. 1984, P.A.S.P. 96, 148
\medskip

\noindent  Arp H. 1990, J. Astrophys. Astr. (India) 11, 411 
\medskip

\noindent Arp, H. 1997, A\&A 319, 33
\medskip

\noindent Arp, H. 1998a, {\it Seeing Red: Redshifts, Cosmology and Academic
Science}, Apeiron, Montreal
\medskip

\noindent Arp. H, 1998b ApJ 496, 661
\medskip

\noindent Arp, H. 1999, ApJ 525, 594
\medskip

\noindent Arp, H. 2001, ApJ 549, 780
\medskip

\noindent Arp, H. and Russell, D. 2001, ApJ 549, 802
\medskip

\noindent Arp, H., Burbidge, E., Chu, Y., Zhu, X. 2001, ApJL 553 in press and astro-ph/0101538
\medskip

\noindent Brandt, W., Hornschmeier A., Schneider, D., et al 2000, AJ 119, 2349
\medskip

\noindent Burbidge, G. and Napier W. 2001, AJ, 121,21
\medskip

\noindent Grandi, S., B\"ohringer, H., Guzzo, L. et al. 1999, ApJ 514, 148  
\medskip

\noindent da Costa, L., Nicolaci, Pellegrini, P. et al. 1988, ApJ 327, 544
\medskip

\noindent Harris, D., Nulsen, P., Ponman,T., et al. 2000, ApJ 530, L81
\medskip

\noindent Hickson, P. 1994, {\it Atlas of Compact Groups of Galaxies}, Gordon and Breach
\medskip

\noindent Knapp,G, Henry, J., Briel, U. 1996, ApJ 472, 125
\medskip

\noindent Kronberg, P., Burbidge, E., Smith, H., Strom, R. 1977, ApJ 218, 8
\medskip

\noindent Lonsdale, C., Hartley-Davies R., Morison, I. 1983, MNRAS 202, 1L
\medskip

\noindent Maia, M., da Costa, L., Latham, D. 1989, ApJS 69 809
\medskip

\noindent Narlikar J., Arp H. 1993, ApJ 405, 51
\medskip

\noindent Ramella, M., Focardi, P., Geller, M. 1996, A\&A 312, 745
\medskip

\noindent Rottgering, H., Wieringa, M., Hunstead, R., Ekers, R. 1997, MNRAS, 290 577
\medskip

\noindent Saraiva, M., Ferrari, F., Pastoriza, M. 1999, A\&A 350, 399
\medskip

\noindent Vikhlinin A., Markevitch, M., Murray S. 2000, astro-ph/0008496   
\medskip

\noindent Sandage, A. 1999, ARA\&A 37, 445
\medskip

\end{document}